# Artificial Intelligence as a Catalyst for Innovation in Software Engineering

Carlos Alberto Fernández-y-Fernández, *Universidad Tecnológica de la Mixteca* Oaxaca, México; Jorge R. Aguilar Cisneros, *Secretaría de Ciencia, Humanidades, Tecnología e Innovación (SECIHTI-Puebla) & UPAEP* University, Puebla, Pue., México.

*Abstract*—The rapid evolution and inherent complexity of modern software requirements demand highly flexible and responsive development methodologies. While Agile frameworks have become the industry standard for prioritizing iteration, collaboration, and adaptability, software development teams continue to face persistent challenges in managing constantly evolving requirements and maintaining product quality under tight deadlines. This article explores the intersection of Artificial Intelligence (AI) and Software Engineering (SE), to analyze how AI serves as a powerful catalyst for enhancing agility and fostering innovation. The research combines a comprehensive review of existing literature with an empirical study, utilizing a survey directed at Software Engineering professionals to assess the perception, adoption, and impact of AI-driven tools. Key findings reveal that the integration of AI (specifically through Machine Learning (ML) and Natural Language Processing (NLP) )facilitates the automation of tedious tasks, from requirement management to code generation and testing . This paper demonstrates that AI not only optimizes current Agile practices but also introduces new capabilities essential for sustaining quality, speed, and innovation in the future landscape of software development.

## I. Introduction

I N today's fast-paced digital world, software engineering is at the heart of technological innovation, driving progress across industries. However, the complexity and constant change of software requirements have made it increasingly difficult for traditional, rigid development methodologies to keep pace. This growing need for flexibility and responsiveness gave rise to the concept of "agility" in software development –Agility is a team's ability to rapidly and reliably respond to change–. Agile methodologies, which prioritize iterative progress, collaboration, and adaptability, have become essential in addressing the ever-shifting demands of modern software systems. Despite the widespread adoption of agile frameworks, software development teams still face significant challenges, such as managing evolving requirements, integrating diverse teams, and ensuring product quality under tight deadlines.

Enter Artificial Intelligence (AI), a transformative technology poised to redefine the landscape of software engineering. AI's ability to automate tasks, process vast amounts of data, and make informed decisions in real time offers a unique opportunity to enhance agility in software development. By leveraging AI-driven tools, teams can streamline workflows, improve accuracy, and innovate at an unprecedented rate. This article delves into the intersection of AI and agile software engineering, exploring how AI not only addresses the traditional challenges of agility but also acts as a catalyst for innovation, enabling more dynamic, efficient, and creative software solutions.

In software engineering, the shift toward agility in development has been a critical response to longstanding challenges such as meeting tight deadlines, changing requirements, and maintaining quality. Traditional, plan-based methods (characterized by structured and linear processes) often struggled to adapt to the dynamic nature of software projects, leading to delays, budget overruns, and the inability to address customer needs adequately [1]. This difficulty, often referred to as the "software crisis," prompted the development of agile methodologies, which emphasize flexibility, collaboration, and responsiveness to change [2].

However, the adoption of agility introduces its own set of challenges. For instance, while agile methodologies facilitate faster development cycles and adaptability, they may also face issues such as integrating multidisciplinary teams and aligning

¹This paragraph of the first footnote will contain the date on which you submitted your paper for review, which is populated by IEEE. It is IEEE style to display support information, including sponsor and financial support acknowledgment, here and not in an acknowledgment section at the end of the article. For example, "This work was supported in part by the U.S. Department of Commerce under Grant 123456." The name of the corresponding author appears after the financial information, e.g. (Corresponding author: Second B. Author). Here you may also indicate if authors contributed equally or if there are co-first authors.

The next few paragraphs should contain the authors' current affiliations, including current address and e-mail. For example, First A. Author is with the National Institute of Standards and Technology, Boulder, CO 80305 USA (e-mail: author@ boulder.nist.gov).

Second B. Author Jr. was with Rice University, Houston, TX 77005 USA. He is now with the Department of Physics, Colorado State University, Fort Collins, CO 80523 USA (e-mail: author@lamar.colostate.edu).

Third C. Author is with the Electrical Engineering Department, University of Colorado, Boulder, CO 80309 USA, on leave from the National Research Institute for Metals, Tsukuba 305-0047, Japan (e-mail: author@nrim.go.jp).

Mentions of supplemental materials and animal/human rights statements can be included here.

Color versions of one or more of the figures in this article are available online at http://ieeexplore.ieee.org

project management processes in large-scale projects [3]. Moreover, organizations often struggle with balancing the formal structure of traditional methods with the iterative, flexible nature of agile development, leading to partial implementations and inconsistent results [4].

AI plays a transformative role in modern technology, impacting various sectors including healthcare, finance, education, and manufacturing. AI's ability to mimic human cognition, such as learning, reasoning, and decision-making, allows for unprecedented advancements across industries. For instance, in healthcare, AI facilitates more accurate diagnostics and treatment recommendations, while in finance, it enhances fraud detection and predictive analytics [5].

The advent of machine learning and deep learning has pushed AI beyond traditional automation, enabling systems to process large volumes of data, recognize complex patterns, and make real-time decisions with remarkable accuracy. These technologies, such as neural networks, underpin many of today's AI applications, including self-driving cars, virtual assistants, and advanced robotics [6]. Moreover, AI's integration into the Internet of Things (IoT) and cloud computing has amplified its reach, allowing for enhanced connectivity and intelligence in various devices and systems [7].

Despite its benefits, AI also presents challenges, including issues related to privacy, bias, and transparency. These challenges need to be addressed to fully harness AI's potential and ensure its ethical application [8].

The objective of this article is to explore how AI enhances agility and fosters innovation within Software Engineering (SE). AI has the potential to optimize various stages of the software development lifecycle by automating repetitive tasks, such as coding, bug detection, and testing, allowing developers to focus on higher-level problem-solving and creativity [9]. Moreover, AI-driven tools can streamline agile development processes by enabling more efficient project management, real-time collaboration, and predictive analytics, which enhance the team's ability to adapt to changing requirements [10].

AI can also act as a catalyst for innovation by facilitating the development of more sophisticated software products. By leveraging machine learning and data-driven insights, AI helps to identify patterns in user behavior, predict market needs, and accelerate the release of innovative solutions that are more tailored to end-user demands [11]. Furthermore, AI contributes to strategic agility in the face of market turbulence, helping firms to innovate and stay competitive by providing real-time data insights and optimizing decision-making processes [12].

In summary, this article aims to highlight the significant role of AI in enhancing agility and fostering innovation in software engineering by transforming workflows, enabling predictive insights, and automating processes, ultimately allowing teams to build more adaptive and creative solutions. The next section details the related work; section III explains how AI is driving agility in SE; section IVdescribes the survey we applied of the use of AI on SE; section V discusses the results of the study; based on the survey; section VI addresses the challenges and limitations associated with using AI in SE; and finally, sections VII suggests areas for future work.

## II. Related Work

### A. Previous research on Agile methodologies and their evolution.

The evolution of Agile methodologies has been driven by the need to improve software development processes and address limitations in traditional approaches. Agile emerged in the 1990s as a response to the rigidity of plan-driven models like Waterfall, which often struggled to accommodate changing requirements and rapid development cycles. The Agile Manifesto, published in 2001, emphasized key values such as individuals and interactions, customer collaboration, and adaptability over processes and tools [17].

Since its introduction, Agile methodologies such as Scrum, XP, and Kanban have been widely adopted and adapted across industries. Research highlights that Agile's flexibility makes it well-suited for dynamic environments, where customer needs and project scope often evolve during the development process [18]. Over time, Agile has evolved into hybrid models like DevOps and Continuous Integration/Continuous Deployment (CI/CD), which further streamline the development cycle by integrating testing, deployment, and operations into the Agile framework [19].

Agile's adaptability has made it a key methodology in modern software development, but its implementation comes with challenges, especially in large-scale projects and distributed teams. Studies have identified factors like organizational culture, team structure, and customer involvement as critical for the success of Agile projects [20]. The evolution of Agile continues as organizations experiment with and refine its methodologies to better fit their specific contexts.

This body of research highlights Agile's significant impact on software engineering, illustrating both its strengths in promoting flexibility and its evolving challenges in complex environments.

### B. The intersection of AI and software engineering.

The intersection of AI and software engineering has gained traction as AI technologies are increasingly integrated into the software development lifecycle. Recent research highlights how AI-driven techniques, such as machine learning (ML), natural language processing (NLP), and deep learning (DL), contribute to more efficient software engineering practices. AI has been applied to automate repetitive tasks, improve testing processes, and enhance decision-making capabilities within software development teams [13].

Several studies have explored the transformative potential of AI in software engineering, noting that AI-based tools can streamline processes such as debugging, testing, and documentation, leading to significant reductions in time and cost [11]. By automating traditionally labor-intensive processes, AI empowers developers to focus on more complex, creative tasks, further accelerating innovation in software products. Moreover, research indicates that AI enables better analysis of large datasets, identifying patterns that can lead to more adaptive and responsive software systems [14].

As AI continues to evolve, it is expected to further reshape software engineering methodologies, with studies advocating for the development of AI-specific frameworks and tools to fully leverage its potential in solving complex software challenges [15], [16]. However, challenges such as model interpretability, ethical concerns, and integration complexities remain areas for further research and improvement [11].

This growing body of literature underscores the critical role AI plays in enhancing the agility and innovation potential of software engineering, laying the foundation for future advancements in both fields.

*C. AI tools and frameworks used in software development.*

In recent years, AI tools and frameworks, particularly Large Language Models (LLMs), have made significant contributions to software development by automating various tasks, improving efficiency, and enhancing coding practices. One prominent example is Copilot (developed by GitHub and OpenAI), which assists developers by generating code suggestions, debugging, and even proposing entire code blocks during development. Studies have shown that tools like Copilot are becoming widely used in the industry, with developers adopting these AI-powered solutions to expedite coding processes and improve code quality [21].

LLMs such as GPT-4 have also demonstrated capabilities in automating more complex software development tasks, including embedded system development and debugging. Research on GPT-4 shows its capacity to generate functional programs for tasks requiring a combination of hardware and software knowledge. This has paved the way for integrating LLMs into workflows, enabling more efficient and accurate software development in specialized areas like embedded systems [22].

Moreover, LangChain, an open-source framework, has facilitated the rapid development of LLM-based applications by providing modular abstractions that allow seamless interaction with various data sources and programming environments. This tool has made it easier for developers to integrate AI-driven functionalities into their software, further enhancing productivity and creativity in development workflows [23].

Additionally, AI tools such as CodeCompose, an AI-assisted code authoring tool developed at Meta, leverage LLMs to improve the speed and accuracy of code generation across multiple programming languages. CodeCompose has been deployed at scale, demonstrating significant improvements in code suggestion acceptance rates and overall development efficiency [24]. LLMs and AI frameworks such as Copilot, GPT-4, LangChain, and CodeCompose are transforming the software development landscape by automating repetitive tasks, generating code, and enhancing productivity, signaling a new era of AI-supported software engineering.

III. AI-DRIVEN AGILITY IN SOFTWARE ENGINEERING

AI has emerged as a transformative force in software engineering, significantly enhancing agility by streamlining processes, improving decision-making, and enabling predictive insights. By automating repetitive tasks and optimizing resource allocation, AI fosters responsiveness and flexibility throughout the software development life cycle (SDLC). The integration of ML models and LLMs has further strengthened this agility by accelerating software delivery, enhancing quality assurance, and improving project management.

*A. AI in Requirements Engineering*

AI techniques, particularly NLP and ML, are reshaping how requirements are elicited, analyzed, and managed. NLP-based tools assist in clarifying vague requirements, extracting structured information from natural language input, and converting it into Agile-aligned user stories, thereby streamlining backlog grooming and sprint planning [25].

Recent work has explored AI for improving user story quality evaluation [26], automatic generation of user stories [27], and intelligent support for requirement engineering [28]. Furthermore, ML-based approaches have been applied to improve requirement engineering practices by detecting inconsistencies and estimating change impact [29].

These advances make requirement processes more consistent, traceable, and adaptive to changing user needs—key characteristics of agile methodologies.

*B. AI in Design and Planning*

AI-driven predictive analytics enable software teams to anticipate risks such as technical debt, performance bottlenecks, and project delays. By analyzing historical project data, ML models can forecast potential issues before they manifest, allowing proactive mitigation strategies that improve delivery timelines and project outcomes [16], [30].

In Agile contexts, predictive analytics support sprint planning and backlog prioritization by identifying high-value features and resource constraints [31], [32]. Bayesian models have even been applied to automate backlog sorting based on feature usage data and success rates, enhancing decision-making and reducing bias [33].

AI-based decision support systems also help project managers balance trade-offs between cost, scope, and time, ensuring optimal resource allocation and reduced risk [34]. These contributions reinforce agility through informed, data-driven planning.

*C. AI & Coding*

Software construction is among the most transformed areas by AI. LLMs such as Codex, AlphaCode, and GitHub Copilot assist developers by generating boilerplate code, refactoring functions, completing code, and creating documentation [35], [36], [37]. These tools reduce development time and human error while enabling developers to focus on high-impact architectural and design decisions.

Other AI-based systems, such as CodeWhisperer [38], extend these capabilities by providing secure code suggestions and context-aware assistance. AI-powered pair programming solutions further increase productivity and collaboration between human developers and intelligent agents [39].

*D. AI in Testing and Quality Assurance*

Testing and quality assurance (QA) have seen significant improvements through AI-driven automation. ML-based predictive models identify defect-prone areas, prioritize test cases, and detect performance issues early [40], [41], [42].

AI frameworks automate test case generation, anomaly detection, and defect prediction using historical data [43]. This reduces manual testing effort and enables continuous integration and delivery [44].

Additionally, AI automates code reviews and bug detection with accuracy levels reaching up to 95%, improving consistency and allowing developers to focus on problem-solving rather than repetitive tasks [35], [45]. These capabilities directly support agile practices by shortening feedback loops and enabling faster, more reliable releases.

*E. AI in Maintenance and Operations*

AI also supports agility beyond deployment by improving software maintenance and operational monitoring. Deep learning–based tools can detect, classify, and repair defects automatically, significantly reducing maintenance costs [46].

Anomaly detection systems proactively monitor performance and identify irregularities in production environments, alerting teams before failures occur [40]. AI-driven automation also assists in continuous deployment and bug fixing, ensuring system stability and reliability throughout its lifecycle [47].

*F. AI in Project Management and Team Collaboration*

In project management, AI facilitates data-driven decision-making and continuous improvement. By analyzing large volumes of project data, AI models forecast delivery times, detect resource constraints, and recommend optimal schedules [30].

NLP and LLM-based assistants such as GPT models enhance team collaboration by summarizing documentation, suggesting improvements, and supporting natural language interaction [37], [48]. These tools also help automate communication tasks and documentation updates, maintaining alignment between teams and stakeholders.

Across all phases of the software life cycle (requirements, design, development, testing, maintenance, and project management), AI acts as a catalyst for agility. It accelerates coding, improves decision-making, reduces defects, and enables proactive maintenance. The integration of ML and LLM technologies transforms software engineering into a more adaptive, efficient, and innovation-driven discipline [49], [50], [51]. These AI-enabled advancements redefine agility not merely as faster delivery, but as an intelligent, data-informed process capable of continuous learning and improvement.

## IV. SURVEY OF AI IN SE

*A. Methodology*

*Survey Design and Distribution*

To investigate perceptions of AI in SE, we designed and distributed an online survey, targeting both academics and industry professionals. The primary goal was to capture current experiences with AI tools, their perceived impact on agility and innovation, and expectations for future developments in AI-assisted software engineering.

The survey was structured into five main sections: (1) demographics and professional background, (2) current AI adoption in development, (3) impact on agility and development processes, (4) innovation and creativity, and (5) challenges, concerns, and future perspectives. A set of optional open-ended questions was also included to gather qualitative insights. The full list of survey questions is provided in Appendix A.

The instrument was distributed via email to over 1,000 addresses of researchers, practitioners, and professionals in the field of software engineering. Participation was voluntary, anonymous, and no identifying information was collected. Respondents were informed that the data would be used solely for research purposes.

A total of 64 complete responses were collected. While the response rate was modest, the data provides valuable exploratory insights into the state of AI adoption and its influence on agility and innovation in software engineering practices. Given the scope of this research, the results are interpreted in a descriptive and exploratory manner rather than aiming for statistical generalization.

*Data Analysis*

The responses were analyzed using both quantitative and qualitative approaches.

- **Quantitative analysis:** Closed-ended questions were summarized using descriptive statistics, including frequency distributions and percentages. These measures provide an overview of respondents' roles, years of experience, development methodologies, AI adoption rates, and perceptions of impact on agility, productivity, and innovation.

- **Qualitative analysis:** Open-ended responses were reviewed and coded thematically to identify recurring patterns, concerns, and perspectives. This analysis highlights nuanced insights into specific use cases of AI in software engineering, perceived limitations, and anticipated future capabilities.

The combination of descriptive statistics and thematic coding provides a balanced view of how AI is currently being integrated into software engineering practices and how it is expected to evolve.

*B. Instrument Validation*

In this paper, a survey was applied, and to evaluate its reliability, an Exploratory Factor Analysis was conducted. In this sense, some constructs were defined.

Construct 1 (C1): AI usage & adoption readiness. This construct measures how embedded AI is in daily work and organizational preparedness. Q5 and Q15 were included: Q5 How frequently do you use AI tools in your daily development work? Q15 Rate your organization's readiness to adopt AI-driven development practices.

Construct 2 (C2): Perceived impact of AI on Agility & Productivity. This construct measures operational benefits. Q6 and Q8 were included: Q6 To what extent has AI improved your team's agility in responding to changing requirements? Q8. How has AI affected your development speed and productivity?

Construct 3 (C3): Perceived impact of AI on Creativity & Innovation. This construct measures higher-order cognitive impact. Q9 and Q10 were included: Q9 Do you believe AI tools enhance or limit creative problem-solving in software development? Q10 Has AI helped your team develop more innovative software solutions?

Before EFA was applied, the adequacy of the data for factor analysis was assessed. In order to do that, the Kaiser–Meyer–Olkin (KMO) index was computed. This index was 0.870, indicating a very good level of shared variance among the items ( Q5, Q6, Q8, Q9, Q10, Q13, and Q15), see Table 1.

**Table 1.** Kaiser-Meyer-Olkin Test

| Items | MSA |
|---|---|
| Q10 Has AI helped your team develop more innovative software solutions? | 0.857 |
| Q13 In the next 5 years, how do you expect AI to transform software engineering? | 0.848 |
| Q15 Rate your organization's readiness to adopt AI-driven development practices: | 0.889 |
| Q5. How frequently do you use AI tools in your 6 development work? | 0.833 |
| Q6 To what extent has AI improved your team's agility in responding to changing requirements? | 0.882 |
| Q8 How has AI affected your development speed and productivity? | 0.909 |
| Q9 Do you believe AI tools enhance or limit creative problem-solving in software development? | 0.862 |
| Overall MSA | 0.870 |

Furthermore, Bartlett's Test of Sphericity was computed to determine if the dataset's correlation matrix significantly differs from an identity matrix. The result is statistically significant ($\chi^2 = 197.802$, df = 21, p < .001), confirming that the correlation matrix is not an identity matrix. These results demonstrate that the data are suitable for factor analysis. At this point, an Exploratory Factor Analysis (EFA) was conducted to examine data from Q5, Q6, Q8, Q9, Q10, Q13, and Q15 to evaluate perceptions of Artificial Intelligence (AI) use and impact in software engineering.

Factor extraction using principal axis factoring with Varimax rotation revealed a single-factor solution. The adequacy of this solution was supported by a non-significant model chi-square ($X^2 = 18.688, df = 14, p = 0.177$) indicating that the one-factor model adequately fits the observed data.

Items related to innovation (Q10), productivity (Q8), and Agility (Q6). Demonstrated the strongest loadings, suggesting that these aspects are central to the construct being measured. Items reflecting Frequency of AI usage (Q5), organizational readiness (Q15), future impact (Q13), and creative problem-solving (Q9) also showed meaningful contributions to the factor (See Table 2). All items loaded positively and substantially on the extracted factor, with standardized loadings ranging from 0.58 to 0.86, exceeding the recommended minimum threshold of 0.50.

**Table 2.** Factor loadings.

| Items | Factor |
|---|---|
| Q10 Has AI helped your team develop more innovative software solutions? | 0.855 |
| Q8 How has AI affected your development speed and productivity? | 0.800 |
| Q6 To what extent has AI improved your team's agility in responding to changing requirements? | 0.774 |
| Q5. How frequently do you use AI tools in your development work? | 0.663 |
| Q15 Rate your organization's readiness to adopt AI-driven development practices: | 0.639 |
| Q13 In the next 5 years, how do you expect AI to transform software engineering? | 0.633 |
| Q9 Do you believe AI tools enhance or limit creative problem-solving in software development? | 0.582 |

*Note.* Varimax was the rotation method applied.

The results support a unidimensional structure for the scale. The purpose of applying EFA was to explore the latent structure underlying perceptions of AI adoption and impact in software engineering. The findings provide strong empirical support for a single latent construct, which can be interpreted as perceived impact and readiness of AI in software engineering practices. This study shows that respondents perceive AI adoption not as a set of isolated dimensions, but rather as a holistic phenomenon encompassing operational, strategic, and cognitive aspects. The high loadings associated with productivity, agility, and innovation indicate that practitioners primarily associate AI value with tangible performance improvements and adaptive capacity. Moderate loadings for items related to frequency of AI use (Q5) and organizational readiness (Q15) suggest that behavioral adoption and strategic preparedness are important, though secondary. Perceptions of creative problem-solving (Q9) indicate that AI is viewed not only as an automation tool but also as a cognitive aid in software development activities.

The results demonstrate satisfactory construct validity at the exploratory level and support the use of a single composite score in subsequent analysis.

An additional test was applied in order to assess the internal consistency and reliability of the proposed scale measuring perceptions of AI impact and adoption in software engineering. An unidimensional reliability analysis was conducted through McDonald's Omega (ω) and Cronbach's Alpha (α). Both coefficients were computed, ω was 0.882, whit 95% confidence interval ranging from 0.837 to 0.926, suggesting strong reliability of the latent construct; α was 0.872, whit 95% confidence interval ranging from 0.807 to 0.937, showing the same situation, a strong reliability, see Table 3. These results indicate high internal consistency.

**Table 3.** Reliability Statistics.

| Coefficient | Estimate | Std. Error | 95% CI Lower | 95% CI Upper |
|---|---|---|---|---|
| Coefficient ω | 0.882 | 0.023 | 0.837 | 0.926 |
| Coefficient α | 0.872 | 0.033 | 0.807 | 0.937 |

In other words, the outcomes confirm that the items consistently measure the same underlying construct.
Additionally, each item was examined using item-rest correlation. As a result, all items exhibited positive and substantial item-rest correlations, ranging from 0.538 to 0.791. See Table 4.

**Table 4.** Item-Rest Correlation and 95% Confidence Intervals for Survey Items.

| Item | Estimate | Lower 95% CI | Upper 95% CI |
|---|---|---|---|
| Q6 To what extent has AI improved your team's agility in responding to changing requirements? | 0.722 | 0.579 | 0.822 |
| Q8 How has AI affected your development speed and productivity? | 0.738 | 0.601 | 0.833 |
| Q9 Do you believe AI tools enhance or limit creative problem-solving in software development? | 0.538 | 0.336 | 0.692 |
| Q10 Has AI helped your team develop more innovative software solutions? | 0.791 | 0.676 | 0.868 |
| Q13 In the next 5 years, how do you expect AI to transform software engineering? | 0.594 | 0.408 | 0.733 |
| Q15 Rate your organization's readiness to adopt AI-driven development practices: | 0.592 | 0.405 | 0.731 |
| Q5. How frequently do you use AI tools in your development work? | 0.619 | 0.44 | 0.75 |

The outcomes indicate that each item contributes meaningfully to the overall construct and that none of them show problematic misalignment.

*C. Results of the survey*

This section presents the survey findings regarding Artificial Intelligence role plays in software engineering. The analysis is based on responses from 64 participants representing both academic and industry perspectives. Results are organized according to the structure of the survey: demographics and professional background, current AI adoption, impact on agility and development processes, contributions to innovation and creativity, as well as challenges, concerns, and expectations for the future.

The aim of this section is not only to provide descriptive statistics of the responses but also to highlight patterns and insights that may inform a broader understanding of how AI is currently influencing software engineering practices.

*Demographics & Background*

Primary Role of Respondents:

The survey participants represented a mix of professional and academic backgrounds, with a clear predominance of research-oriented roles. Out of the 64 respondents, approximately two-thirds identified as Research Scientists or Academic Researchers, reflecting strong engagement from the academic community.

The second largest group consisted of Software Engineers/Developers, who accounted for a significant share of the responses and provided insights from an industry practice perspective. Smaller groups were represented by Engineering Managers/Team Leads, Quality Assurance Engineers, Product Managers, and teaching professionals (e.g., professors or instructors).

This distribution indicates that the results primarily reflect perspectives from academia, complemented by contributions from practitioners in software engineering and related roles. The combination of both viewpoints enriches the analysis by integrating theoretical and practical experiences with AI in software engineering, see Fig. 1.

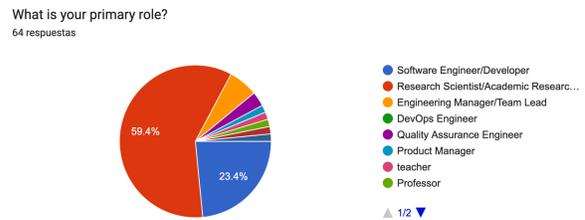

**Fig. 1.** Primary Role of Respondents.

Years of Experience in Software Development:

The respondents reported a wide range of professional experience, with a notable concentration among highly experienced professionals. The largest group consisted of participants with more than 16 years of experience, representing nearly half of all responses. This indicates that the survey captured the perspectives of individuals with long-standing involvement in software engineering practices and their evolution over time.

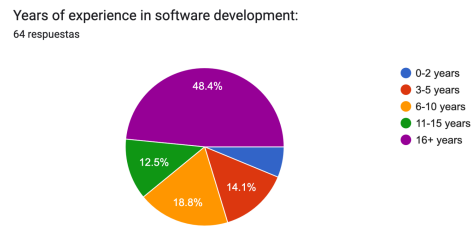

**Fig. 2.** Years of Experience in Software Development.

The second most represented categories were professionals with 6–10 years and 11–15 years of experience, both of which together accounted for a substantial portion of the sample. These groups provide insights from mid-career practitioners and researchers actively engaged in current development practices. Smaller groups were reported at the 3–5 years and 0–2 years levels, showing that while early-career voices were present, they were less prominent in the dataset.

Overall, the distribution suggests that the findings reflect the perspectives of a predominantly experienced population, which may influence how AI adoption, agility, and innovation are perceived in the context of software engineering, see Fig. 2.

Development Methodologies in Use:

The survey responses show a strong prevalence of Agile/Scrum practices among participants. A clear majority reported using Agile/Scrum either as their primary methodology or in combination with other approaches. This reflects a widespread adoption of agile methods as a standard in both academic and industrial contexts of software engineering.

In addition to Agile/Scrum, many respondents indicated the use of hybrid approaches, combining agile practices with more traditional methodologies such as Waterfall or Rational Unified Process (RUP). This suggests that while agile dominates, organizations often adapt methodologies to fit specific project or organizational needs rather than applying a single model uniformly.

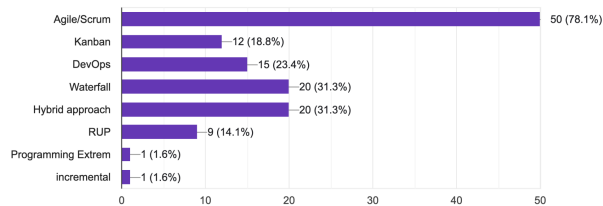

**Fig. 3.** Development Methodologies in Use.

DevOps also appeared as a frequently mentioned methodology, typically in combination with Agile/Scrum. This pairing highlights the growing importance of continuous integration, delivery, and operations as integral parts of software development. Kanban was reported less frequently but was present in several responses, often combined with Scrum or Waterfall.

A smaller number of participants identified Waterfall or RUP as standalone or dominant methodologies, although these were generally reported as part of mixed or hybrid models. This indicates that while traditional models are not entirely absent, they are less prominent compared to agile-oriented frameworks.

The results suggest that agile practices—especially Scrum—serve as the foundation of most teams' development methodologies, often complemented by hybridization with DevOps, Kanban, or legacy models such as Waterfall and RUP. This reflects a trend toward flexibility and methodological integration rather than strict adherence to a single process model, see Fig. 3[2].

*Current AI Adoption in Development*

AI-Powered Tools in Development Workflows:

The responses indicate that the most widely adopted AI-powered tools are GitHub Copilot and ChatGPT/Gemini/GPT-based coding assistants, with the majority of participants reporting the use of one or both in their development activities. These tools appear to dominate current AI adoption in software engineering, reflecting their accessibility, integration into coding workflows, and immediate impact on productivity.

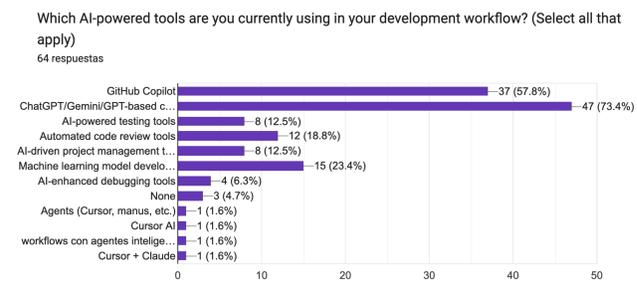

**Fig. 4.** AI-Powered Tools in Development Workflows.

Beyond these two leading categories, several respondents reported the use of machine learning model development platforms, AI-powered testing tools, and automated code review tools. These applications suggest that AI is being integrated not only in code generation but also in quality assurance and specialized areas such as debugging and project management. A few participants mentioned the use of emerging tools like Cursor AI and intelligent agents, highlighting experimentation with newer platforms.

A small proportion of respondents indicated no current use of AI tools, suggesting that while adoption is widespread, it is not yet universal.

In general, the data demonstrates that AI integration in software engineering is strongly concentrated around code assistance and review tools, with gradual but growing adoption of AI in complementary areas such as testing, debugging, and project management. This pattern suggests that AI's most immediate value is perceived in coding-related tasks, while its application in broader lifecycle activities is still developing, see Fig. 4.

Frequency of AI Tool Usage:

---

[2] Note: As participants could select multiple options for some questions, the total percentage exceeds 100%.

The results reveal that AI tools have already become part of the regular workflow for most respondents. A large proportion reported using them daily or several times per week, together representing the majority of the sample. This indicates that for many professionals and researchers, AI has moved beyond experimental use and has been integrated into routine development practices.

How frequently do you use AI tools in your daily development work?
64 respuestas

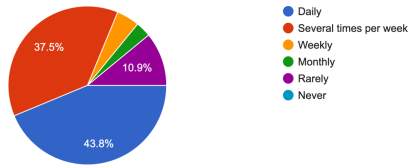

**Fig. 5.** Frequency of AI Tool Usage.

A smaller group reported using AI tools on a weekly or monthly basis, suggesting more occasional adoption, often for specific tasks rather than continuous support. Finally, only a limited number of participants indicated using AI tools rarely, highlighting that outright non-use is relatively uncommon among respondents, see Fig. 5.

The findings suggest that AI is not only widely adopted but also frequently used, particularly for day-to-day coding, debugging, and related activities. This level of regular engagement underscores the growing reliance on AI-powered tools in software engineering practice.

*Impact on Agility & Development Process*

Impact of AI on Teams' Agility in Responding to Changing Requirements:

The survey responses indicate that AI has had a predominantly positive impact on teams' agility when adapting to changing requirements. A substantial portion of participants reported that AI significantly improved their ability to respond, highlighting a strong perception of value in this area. Another large segment indicated that agility was moderately improved, suggesting that while benefits are noticeable, they may vary depending on context, adoption level, or integration with existing practices. A smaller group noted only slight improvements or no change, pointing to possible limitations in applicability or effectiveness. Finally, a minority of respondents stated that the question was not applicable to them, as they do not currently use AI tools in their development process, see Fig. 6.

To what extent has AI improved your team's agility in responding to changing requirements?
64 respuestas

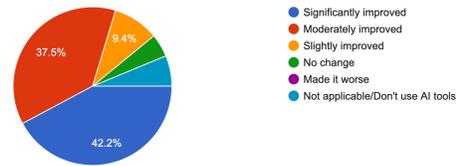

**Fig. 6.** Impact of AI on Teams' Agility in Responding to Changing Requirements.

Together, these findings suggest that AI is generally perceived as a facilitator of adaptability, with most teams experiencing tangible gains in flexibility, speed, and responsiveness to evolving requirements. However, the mixed responses also underline that the degree of improvement depends on how extensively AI is integrated into workflows and supported by organizational practices.

Stages of the Development Lifecycle Where AI Is Most Beneficial:

The survey results show that AI is perceived as most beneficial in the coding/implementation stage. A large majority of respondents identified this phase as the primary area where AI tools add value, reflecting their growing use in code generation, refactoring, and error detection.

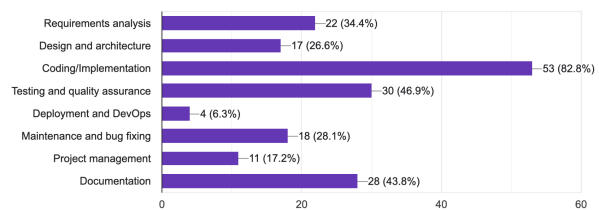

**Fig. 7.** Stages of the Development Lifecycle Where AI Is Most Beneficial.

The second most frequently cited stage was testing and quality assurance, where AI supports activities such as automated test generation, bug detection, and performance validation. This finding highlights the role of AI in enhancing software reliability and reducing the effort required in manual testing.

Other lifecycle stages received fewer mentions but still demonstrated meaningful adoption. These included maintenance and bug fixing, and documentation, both of which benefit from AI-assisted error resolution and the automation of repetitive documentation tasks. A smaller but notable share of participants pointed to requirements analysis and design and architecture, suggesting an emerging role for

AI in upstream activities such as requirement validation, early modeling, and system design support. Less frequently, AI was associated with project management and deployment/DevOps, indicating more limited yet developing areas of application, see Fig. 7.

Generally speaking, the data reveals a clear trend: AI is most valued in stages involving code production and validation, while its use in earlier and managerial phases of the lifecycle remains exploratory but promising. This suggests that adoption is currently concentrated in technically intensive activities, with potential expansion toward more strategic stages as tools mature.

Impact of AI on Development Speed and Productivity:

The survey results indicate that AI has had a substantial impact on developers' speed and productivity. The most common response was "increased significantly (>30%)", reported by a large proportion of participants. This suggests that, for many teams, AI adoption translates into marked efficiency gains in daily development tasks.

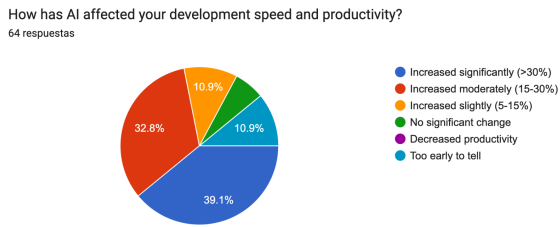

**Fig. 8.** Impact of AI on Development Speed and Productivity.

The second most frequent response was "increased moderately (15–30%)", also widely reported. Taken together, these two categories show that the majority of respondents experienced measurable productivity improvements, ranging from moderate to significant.

A smaller number of participants noted slight increases (5–15%), while a minority reported no significant change in their productivity. Additionally, some respondents selected "too early to tell," reflecting uncertainty or limited exposure to AI tools, see Fig. 8.

The findings show that AI is perceived primarily as a productivity accelerator, with most developers identifying moderate to substantial gains. Cases of no change or uncertainty are exceptions rather than the norm, underscoring the strong positive association between AI adoption and enhanced development performance.

*Innovation & Creativity*

AI and Creative Problem-Solving in Software Development:

The results reveal a predominantly positive perception of AI's role in creative problem-solving. Most respondents indicated that AI enhances creativity, either moderately or significantly. In particular, "significantly enhance creativity" was one of the most frequent responses, suggesting that many practitioners view AI as a catalyst for generating innovative solutions, exploring alternatives, and overcoming technical challenges.

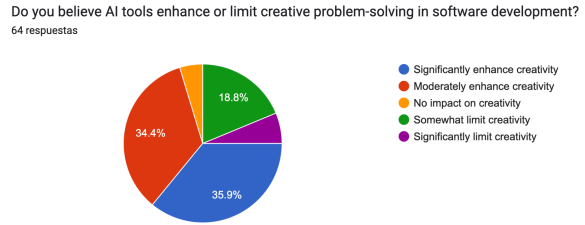

**Fig. 9.** AI and Creative Problem-Solving in Software Development

At the same time, a considerable portion of participants selected "moderately enhance creativity," reflecting a more balanced perspective in which AI is seen as supportive, though not transformative, in fostering creativity.

On the other hand, a smaller but notable group reported that AI tools somewhat or significantly limit creativity. These responses highlight concerns that reliance on AI could discourage original thinking or lead to over-dependence on automated suggestions. A minority of respondents considered that AI had no impact on creativity, indicating neutrality or negligible influence in their workflow.

Taken together, the findings suggest that developers predominantly regard AI as an enabler of creative problem-solving, though a non-negligible segment perceives potential risks of limitation. This divergence underscores the dual role of AI: while it can amplify ideation and experimentation, it may also constrain originality if used uncritically, see Fig. 9.

AI and the Development of Innovative Software Solutions:

The majority of respondents reported that AI has had a positive impact on their team's ability to develop innovative software solutions. A large share selected "Yes, definitely", indicating strong confidence in AI's contribution to fostering new approaches, features, and solution designs. Many others chose "Yes, to some extent," reflecting that AI supports innovation, though its impact may be more incremental or context-dependent.

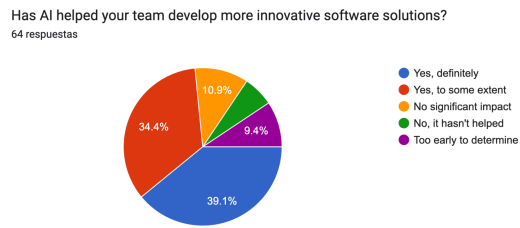

**Fig. 10.** AI and the Development of Innovative Software Solutions.

In contrast, a smaller group of participants indicated "No significant impact" or "No, it hasn't helped," suggesting that AI adoption has not yet translated into greater innovation in their experience. Additionally, some respondents answered "Too early to determine," pointing to the ongoing and exploratory nature of AI integration in development workflows, see Fig. 10.

Overall, these results suggest that AI is widely perceived as an enabler of innovation in software engineering, though its effects are not uniform. While many teams already report tangible benefits, others remain cautious, either due to limited adoption or uncertainty about long-term outcomes.

*Challenges & Concerns*

Concerns About AI Integration in Software Development:

The survey highlights a range of concerns regarding AI adoption in software development, with several recurring themes across respondents. The most frequently mentioned issues were code quality and reliability, security vulnerabilities, and over-dependence on AI tools, underscoring apprehension that automation may introduce undetected errors, weaken security practices, or reduce human oversight.

Privacy and data protection, and the lack of transparency in AI decisions, also emerged as significant points of tension, reflecting worries about compliance, accountability, and the interpretability of AI-generated outputs. Many respondents additionally expressed concern about job displacement, the cost of implementation, and training and skill gaps, signaling broader organizational and workforce challenges beyond the technical dimension.

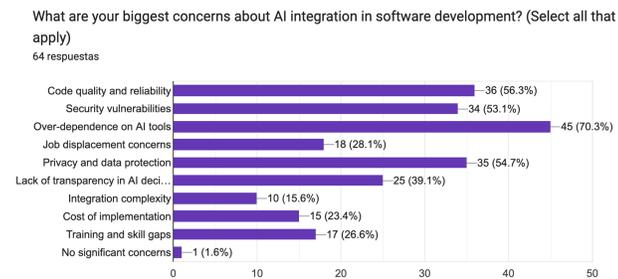

**Fig. 11.** Concerns About AI Integration in Software Development

A smaller set of participants raised issues around integration complexity, pointing to the difficulties of embedding AI systems into existing workflows and toolchains. Notably, only a very limited number indicated having no significant concerns, suggesting that while AI tools are widely used, most practitioners remain cautious about their implications, see Fig. 11.

These results indicate that while AI adoption is advancing, developers and teams remain highly attentive to the risks and trade-offs associated with integration. The findings emphasize the need for governance frameworks, training programs, and transparency measures to ensure that AI tools are deployed responsibly and effectively.

Barriers to Wider AI Adoption in Software Development:

The survey revealed a variety of barriers that hinder the broader adoption of AI in software development processes. Among the most frequently cited challenges were lack of budget and resources, technical integration difficulties, and security and compliance concerns. These findings suggest that both financial and infrastructural constraints remain critical bottlenecks for many organizations.

Another recurring theme was the lack of expertise and training, indicating that teams often feel underprepared to work effectively with AI technologies. This skills gap appears to compound the challenge of integration, making adoption both technically and organizationally demanding. Team resistance to change and the need for management buy-in were also noted, reflecting cultural and organizational hurdles beyond purely technical aspects, see Fig. 12.

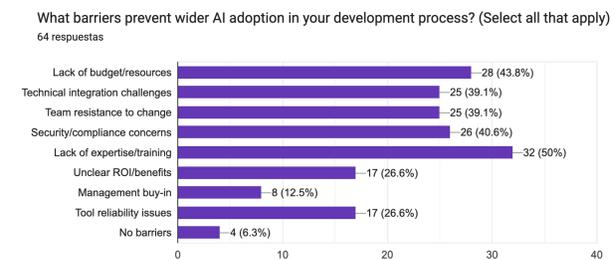

**Fig. 12.** Barriers to Wider AI Adoption in Software Development.

Participants further expressed uncertainty about the return on investment (ROI) and cited tool reliability issues, suggesting that the perceived benefits of AI tools are not always clear or consistent in practice. A smaller but notable subset of respondents indicated that they faced no significant barriers, implying that some teams have already developed the structures and confidence necessary for effective AI integration.

Therefore, the findings show that AI adoption is not solely a question of technical capability but also of resources, expertise, trust, and organizational alignment. Addressing these issues will be essential for fostering wider and more sustainable integration of AI in software engineering practices.

*Future Perspectives*

Expected Transformation of Software Engineering Through AI:

When asked how AI is expected to transform software engineering in the next five years, respondents expressed high expectations of change, though with varying degrees of intensity. The majority anticipated either a significant evolution or a revolutionary transformation. This reflects a strong belief that AI will play a central role in reshaping practices, processes, and tools within the discipline.

A smaller group of participants predicted gradual improvement, suggesting a more incremental trajectory in which AI enhances existing practices rather than fundamentally altering them. Only a minimal number of respondents expected little to no change, reflecting skepticism about AI's disruptive potential. Additionally, a few participants expressed uncertainty, highlighting the unpredictability that still surrounds the pace and direction of AI adoption, see Fig. 13.

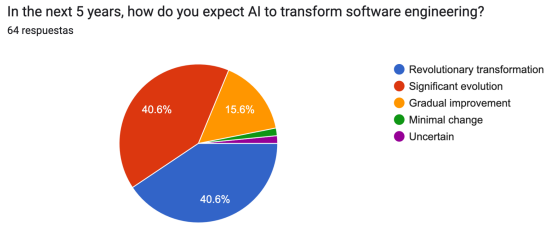

**Fig. 13.** Expected Transformation of Software Engineering Through AI.

The findings show a clear consensus that AI will be transformative for software engineering, though opinions diverge on whether this change will be evolutionary or revolutionary. This distribution suggests both optimism and caution: while most foresee AI as a catalyst for major progress, there is still recognition of possible constraints and uncertainties in its practical realization.

Valuable AI Capabilities for Development Teams:

When asked which AI capabilities would be most valuable for their development teams, respondents consistently highlighted a set of core functionalities that they believe could provide the greatest impact, see Fig. 14.

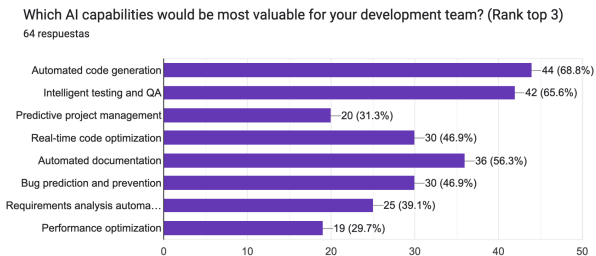

**Fig. 14.** Valuable AI Capabilities for Development Teams.

Automated code generation emerged as the most frequently mentioned capability. Many participants identified it as a way to accelerate development cycles, reduce repetitive coding tasks, and allow developers to focus on higher-level design and problem-solving.

Closely following, intelligent testing and quality assurance (QA) were regarded as essential. Respondents saw this capability as critical for improving software reliability, identifying issues earlier in the lifecycle, and reducing the overall cost of defect management.

Another highly valued set of capabilities included real-time code optimization and automated documentation. These were seen as complementary to code generation and testing, offering efficiency gains in both system performance and maintainability.

Additional capabilities such as bug prediction and prevention, requirements analysis automation, predictive project management, and performance optimization were also frequently mentioned. While these were not ranked as highly as code generation and testing, their repeated appearance indicates that teams view AI as a potential enabler across the entire software engineering pipeline, from early requirements to maintenance.

The results point to a strong expectation that AI will primarily enhance core engineering tasks—coding, testing, and optimization—while also offering secondary benefits in documentation, project management, and predictive analytics. This suggests a dual emphasis: immediate productivity gains and long-term improvements in software quality and sustainability.

Organizational Readiness for AI Adoption:

When assessing their organizations' readiness to adopt AI-driven development practices, responses showed a mixed but telling distribution across different levels of maturity.

As we can see in Fig. 15, the majority of participants reported being "somewhat ready, exploring options." This indicates that many organizations are in an exploratory stage—aware of AI's potential but still evaluating tools, strategies, and feasibility before committing to concrete adoption.

A significant proportion identified as "mostly ready, planning implementation." These organizations have moved beyond exploration and are actively preparing structured plans to integrate AI into their development workflows, signaling a near-term transition to practical deployment.

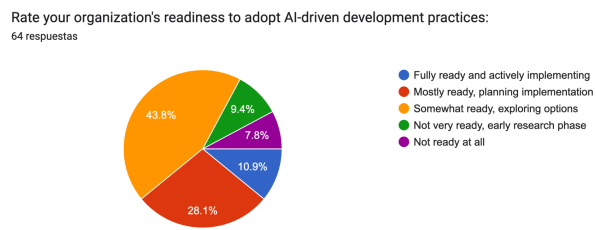

**Fig.** 15. Organizational Readiness for AI Adoption

At the more advanced end, a smaller but notable group reported being "fully ready and actively implementing" AI-driven practices. This demonstrates that while adoption remains uneven, some organizations have already taken decisive steps toward operationalizing AI in software engineering.

Conversely, a minority described themselves as "not very ready, early research phase" or "not ready at all." These cases reflect organizations still at the starting line, either conducting initial investigations or lacking the foundational resources and strategy to engage with AI adoption meaningfully.

In general, the data suggest a readiness landscape skewed toward early- to mid-stage adoption. While enthusiasm and exploration dominate, only a smaller subset has reached active implementation, highlighting both the momentum of interest and the challenges that still slow down full integration.

*Open-Ended Questions*

Examples of AI Impact on Development and Team Agility:

Participants provided diverse examples of how AI has improved their development processes and team agility. A recurring theme was the use of AI-powered code generation. Developers reported that AI tools help produce boilerplate code, accelerate the implementation of specific functions, and even create full-stack applications despite limited knowledge of certain programming languages. This has led to substantial time savings, enabling teams to reallocate effort toward higher-level design and analysis tasks.

Another frequently mentioned area was AI-driven code review and debugging. Several respondents described integrating AI assistants into CI/CD pipelines to detect vulnerabilities, flag logic errors, or generate unit tests. These tools reduced review cycles by 30–40% in some cases and significantly cut debugging times, improving sprint velocity and overall agility. AI-based bug detection and optimization suggestions were also seen as valuable for enhancing code quality and maintainability.

Testing and quality assurance also emerged as a significant domain of impact. Respondents highlighted AI's ability to automatically generate test cases—including edge cases often overlooked by humans—and to support test automation and continuous deployment. This has strengthened defect detection while reducing manual effort.

AI has further contributed to documentation and requirements activities. Teams noted that AI tools accelerate documentation tasks, improve clarity in requirement specifications, and assist with knowledge reuse. Some participants also emphasized AI's role in problem understanding and solution exploration, such as identifying alternative architectures, resolving configuration issues, or proposing optimizations.

Beyond technical coding support, some respondents pointed out broader benefits, such as AI aiding in learning new languages, reducing time spent on troubleshooting, or suggesting new tools and libraries that expand team capabilities.

It is worth noting that not all participants reported positive experiences—several indicated having no significant examples yet, reflecting uneven adoption and maturity across organizations.

The examples illustrate AI's versatility: from automating routine coding and testing tasks to enhancing reviews, documentation, and architectural decisions. The consistent benefits cited—time savings, improved code quality, and greater agility—indicate that AI is already reshaping practical aspects of software development workflows.

Desired but Currently Missing AI Capabilities:

When asked what AI features or capabilities they wished existed but are not currently available, participants provided a wide range of responses that highlight unmet needs and future opportunities.

A central theme was the desire for AI to bridge the gap between requirements and implementation. Several respondents envisioned tools capable of fully understanding natural language business requirements and automatically generating high-quality, maintainable code, complete with tests, documentation, and traceability. Others emphasized the need for AI to handle requirements modeling and evaluation, including automated requirement analysis, verification, and even future modeling based on diverse user profiles.

Another frequent request concerned system-level reasoning and architecture support. Participants noted that current AI excels at localized code generation but lacks the ability to reason about software architecture holistically. Desired features included AI that could proactively suggest refactors, manage technical debt, generate software architecture designs, or recommend appropriate design patterns (e.g., SOLID, DDD, CQRS) at the right stage of development.

On the operational side, developers called for AI integration into DevOps and CI/CD pipelines, with seamless compatibility with tools such as Jenkins and Octopus Deploy. This included features like real-time build support, test automation, deployment optimization, and technical debt prediction across the software lifecycle.

Testing and quality assurance also surfaced as a priority. Participants expressed interest in deterministic testing tools, more reliable test automation, error autocorrection, bug prevention, and features that would allow AI to generate only the part of the code being modified rather than regenerating entire files.

Beyond technical functions, developers voiced concerns about AI usability and trustworthiness. Desired improvements included transparency in responses, consistent code style enforcement, unified context awareness across sessions, explainability of recommendations, and privacy assurances for proprietary code. Some also wished for accuracy indicators, confidence levels, or trust scores to better assess AI outputs.

Finally, several respondents suggested innovative or forward-looking applications. These included model-driven UI generation that integrates usability heuristics, AI-based project management, cost and risk estimation, end-user self-programming, ecological optimizations to reduce carbon footprint, and even applications in embedded systems.

Notably, some participants indicated they were satisfied with current tools or lacked enough experience to identify missing features, underscoring differences in maturity and adoption levels across organizations.

In sum, while current AI tools already provide value in code generation and automation, participants expressed a clear desire for more advanced, explainable, and integrated capabilities that extend across the full software engineering lifecycle, from requirements to deployment and long-term maintenance.

Perspectives on AI's Role in Enhancing Agility and

13Innovation:

Responses to the open-ended question about AI's role in agility and innovation in software engineering revealed a spectrum of views, ranging from enthusiasm and optimism to caution and skepticism.

A large group of participants highlighted the automation of repetitive tasks—such as testing, bug detection, code generation, and documentation—as a primary benefit. By reducing time spent on routine activities, AI is seen as enabling developers to focus more on creative problem-solving, innovation, and decision-making. Several responses framed AI as a true assistant or catalyst for productivity, accelerating development cycles and supporting agile practices through faster context switching, continuous experimentation, and intelligent insights for project management and risk anticipation.

Another recurring theme was the transformative or disruptive nature of AI. Some respondents described it as a "new paradigm" or a "disruptive change" in software engineering, potentially reshaping how teams build software. A few extended this view to predict that AI could eventually generate complete, error-free solutions independently, although most recognized that in the near term, human creativity, reasoning, and lateral thinking remain essential.

Concerns were also evident. Participants pointed to limitations of current AI systems, including their probabilistic nature, lack of transparency, inconsistent outputs, and potential to introduce security or privacy risks. Several highlighted the danger of overreliance on AI, particularly in educational contexts, where dependence could undermine the development of critical thinking and foundational knowledge among students and junior developers. Issues of trust, accuracy, and regulation were raised, with some suggesting the need for clearer standards and safeguards.

Other contributions emphasized integration challenges. Respondents noted that AI is not yet well embedded in existing software engineering methodologies and that further effort is required to harmonize AI capabilities with human expertise across different phases of development. Calls were made for team education, adaptation of curricula, and development of methodologies that integrate generative AI more systematically into agile processes.

Finally, some participants offered forward-looking ideas, such as AI support for requirements analysis, sprint backlog tracking, adaptive planning based on historical team data, and user experience testing. These responses underline the potential of AI to extend its role beyond coding into project management and process optimization.

To summarize, while many respondents see AI as a powerful tool for enhancing agility and innovation, there is broad recognition of its current limitations, the need for cautious integration, and the importance of preserving the uniquely human aspects of creativity and reasoning in software engineering.

## V. Discussion

The survey results provide an exploratory yet informative picture of how AI is currently reshaping software engineering practices. Several key themes emerge when interpreting the findings in light of the broader research objectives.

First, the demographic profile of respondents (dominated by academics and highly experienced professionals) suggests that perspectives captured here are grounded in both long-term expertise and research-oriented observation. This lends weight to the results, though it also introduces a possible bias toward theoretical or experimental viewpoints rather than widespread industry adoption.

Second, the findings confirm that AI integration in software engineering is already substantial, with tools such as GitHub Copilot and GPT-based assistants embedded in daily workflows. This aligns with recent reports of accelerated adoption of AI for code generation and debugging. However, adoption remains uneven, with some respondents still reporting limited or no use. This reflects a dual landscape: on one side, early adopters incorporating AI extensively into coding, testing, and quality assurance; on the other, cautious or resource-limited teams still in exploratory phases.

Third, the reported impact on agility, productivity, and innovation is broadly positive. Most participants identified measurable gains in speed, responsiveness, and creative problem-solving. Importantly, these benefits were not limited to coding tasks but extended to documentation, testing, and, in some cases, requirements engineering. At the same time, concerns about creativity being limited by over-reliance on AI illustrate the nuanced role of automation: while AI can stimulate ideation, it may also constrain originality if used uncritically.

Fourth, challenges and barriers identified highlight that AI adoption is not merely a technical question. Issues such as security, code reliability, skill gaps, and unclear return on investment reveal the organizational and cultural dimensions of AI integration. This indicates that sustainable adoption requires not only technical maturity but also governance frameworks, training, and organizational readiness.

Finally, future perspectives point toward optimism. The majority of respondents expect AI to transform software engineering significantly, either gradually or in a revolutionary manner. The desired capabilities reported (such as requirements-to-code automation, system-level reasoning, and more reliable integration with DevOps) indicate that practitioners envision AI not only as a coding assistant but also as a strategic partner across the entire lifecycle. This expectation highlights the gap between current AI functionality and the aspirational role envisioned by professionals.

Taken together, the discussion suggests three overarching insights. First, AI is already delivering tangible benefits in software engineering, especially in coding and testing. Second, adoption remains constrained by technical, organizational, and cultural barriers that must be addressed for broader uptake. Third, practitioners and researchers alike see AI as a transformative force, though its long-term role will depend on bridging current limitations with the aspirational features identified by respondents.

### A. Limitations of the survey

Some limitations of this study should be acknowledged. The modest sample size constrains the generalizability of findings and means the results should be interpreted as exploratory rather than representative of the entire software engineering community. The respondent pool was heavily skewed toward academics and highly experienced professionals, which may bias the findings toward research-oriented perspectives and underrepresent the views of junior practitioners or industry-only teams. Furthermore, the survey design emphasized descriptive rather than inferential analysis, limiting the ability to draw causal conclusions about the relationship between AI adoption and agility or innovation outcomes. Finally, self-reported data may be subject to overstatement of benefits or understatement of challenges, given the novelty and visibility of AI in the current technological discourse.

## VI. CHALLENGES AND LIMITATIONS

The integration of AI into software engineering presents both opportunities and constraints that must be critically acknowledged. While AI demonstrates clear potential to enhance agility, productivity, and innovation, several challenges remain that could limit its widespread adoption and sustainable impact:

- **Technical limitations.** Current AI systems, particularly LLMs, are prone to inaccuracies, hallucinations, and inconsistent outputs. These issues raise concerns about the reliability of AI-generated code, test cases, and documentation. Moreover, many AI tools still lack the ability to reason about software architecture at a systemic level, restricting their usefulness to localized coding tasks rather than end-to-end engineering processes.
- **Trust, transparency, and accountability.** A recurring challenge is the opacity of AI decision-making. Developers often lack insight into how outputs are generated, making it difficult to assess correctness, trace errors, or ensure compliance with security and regulatory standards. This "black box" effect undermines trust and complicates integration in high-stakes environments where accountability is critical.
- **Organizational and human factors.** Effective adoption of AI requires more than technical readiness. Teams often face resistance to change, skill gaps, and uncertainty about return on investment. Over-dependence on AI tools also raises concerns about the erosion of human expertise, especially in training contexts where critical thinking and foundational software engineering skills are still essential.
- **Ethical and security concerns.** The introduction of AI raises risks related to privacy, intellectual property, and potential bias in generated outputs. Automated code suggestions may inadvertently replicate insecure practices, introduce vulnerabilities, or reuse copyrighted material. These risks demand stronger governance and validation mechanisms before AI-generated artifacts can be deployed at scale.

## VII. FUTURE DIRECTIONS

The findings of our study indicate that AI is emerging as a transformative force in SE, yet its full potential remains only partially realized. Future work should focus on developing integrated AI-driven frameworks that span the entire software development lifecycle. Rather than applying AI in isolated tasks (such as coding, testing, or documentation), we think research should aim to create cohesive architectures that support intelligent collaboration, traceability, and continuous improvement across all development phases.

Another critical direction involves advancing explainability and trust in AI-assisted development. The opacity of current AI models limits their reliability in professional and safety-critical contexts. Future systems should incorporate explainable AI (XAI) [52] capabilities capable of clarifying reasoning processes, indicating confidence levels, and supporting reproducibility. Such transparency would foster greater confidence among developers and facilitate ethical deployment in industrial environments.

The evolving relationship between humans and AI also requires deeper exploration. Research should investigate how intelligent agents can participate as collaborators within agile teams, assisting in planning, retrospectives, and decision-making. This shift from automation to augmentation redefines agility as a joint cognitive process, where human creativity and AI intelligence complement each other.

Equally important is the establishment of ethical and governance frameworks for responsible AI use. As AI systems generate more code and design artifacts, questions of authorship, accountability, data privacy, and bias mitigation must be addressed systematically. The software engineering community will need shared standards to ensure that AI-augmented practices remain transparent, fair, and compliant with regulatory expectations.

Education will play a decisive role in enabling this transition. Curricula in SE should incorporate AI literacy, prompt design, ethics, and collaboration strategies to prepare professionals who can critically integrate AI tools without sacrificing creativity or foundational skills. Finally, empirical and longitudinal research is needed to measure the long-term effects of AI adoption on productivity, quality, and innovation. Systematic experimentation and industrial case studies will provide the evidence base required to formalize best practices in AI-supported software engineering.

We believe that the future of SE lies in harmonizing human expertise with machine intelligence. By building transparent, ethical, and integrated systems, AI can move beyond automation to become a genuine catalyst for creativity, agility, and sustainable innovation in the digital era.

## ACKNOWLEDGMENT

Tx.

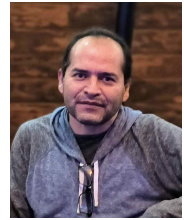

**Carlos Alberto Fernández-y-Fernández received** received the B.S. degree from the Universidad Veracruzana, Mexico, the M.S. degree in computer science from the Fundación Arturo Rosenblueth, Mexico, and the Ph.D. degree in computer science from The University of Sheffield, U.K. His major field of study is software engineering.

He is currently a Full-time Professor and Researcher with the Instituto de Computación, Universidad Tecnológica de la Mixteca (UTM), Oaxaca, Mexico. He served as the Director of the Instituto de Computación, UTM, from 2017 to 2025. Previously, he has coordinated several graduate programs in distributed systems and applied computing. His research interests include visual modeling, agile and traditional development methods, and formal software specification.

Dr. Fernández y Fernández is a member of the Sistema Nacional de Investigadoras e Investigadores (SNII) and the Academia Mexicana de Computación (AMEXCOMP). He is a founding member of the Red Temática Mexicana de Ingeniería de Software (REDMIS).

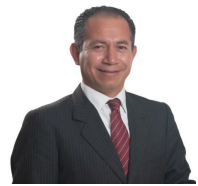

**Jorge Rafael Aguilar Cisneros** was born in México. Dr. Aguilar earned his PhD in Software Engineering at UPAEP. He studied a Master's degree in Systems Engineering at UDALAP, and a Bachelor's degree in Computer Science at BUAP. He is currently a member of the SNII (Mexican National System Researchers).

His research interests focus on the area of Software Engineering. He has written articles for national and international conferences, journals, and book chapters. Dr. Aguilar has been a member of the IEEE, AMEXCOMP (Computing Mexican Academy), and RedMIS. Currently, Dr. Aguilar teaches undergraduate and graduate courses at University UPAEP in México.